\documentclass[10pt, a4paper]{article}

\usepackage[dvipdfmx]{graphicx, color}
\usepackage{wrapfig}
\usepackage{amsmath, amssymb}
\usepackage{bm}
\usepackage{pifont}
\usepackage[dviout]{pict2e}
\usepackage{booktabs}
\usepackage{lscape}
\usepackage{cite}
\usepackage{setspace}

\usepackage{here}

\usepackage{lineno}

\title{Prospects of measuring Higgs boson decays into muon pairs at the ILC}
\author{Shin-ichi Kawada$^{\dagger}$, Jenny List, Mikael Berggren}
\date{}

\begin{document}

\maketitle

\begin{center}
DESY, Notkestra{\ss}e 85, 22607, Hamburg, Germany
\\
$^{\dagger}$ : shin-ichi.kawada@desy.de
\end{center}

\renewcommand{\thefootnote}{\fnsymbol{footnote}}

\begin{abstract}
We study the prospects of measuring the decay of the Higgs boson into a pair of muons at the International Linear Collider (ILC).
The study is performed at center-of-mass energies of 250\,GeV and 500\,GeV, with fully-simulated Monte-Carlo samples based on the International Large Detector (ILD).
The expected precision on cross section times branching ratio $\sigma \times \mathrm{BR}(h \to \mu ^+ \mu ^-)$ has been evaluated to be 24.9{\%} for an integrated luminosity of 2\,ab$^{-1}$ at 250\,GeV.
This result improves to 17.5{\%} in combination with 4\,ab$^{-1}$ of 500\,GeV data.
We also quantify the impact of the transverse momentum resolution on this analysis, and found that it is very important reach the design goal of an asymptotic resolution of $\sigma_{1/P_t} = 2 \times 10^{-5}$\,GeV$^{-1}$.

Talk presented at the International Workshop on Future Linear Colliders (LCWS2018), Arlington, Texas, 22-26 October 2018. C18-10-22.
\end{abstract}

\renewcommand{\thefootnote}{\arabic{footnote}}

\section{Introduction}
After the discovery of the Higgs boson, the detailed investigation of its properties is one of the most important topics of particle physics.
In this contribution, we focus on the measurement the decay of the Higgs boson into a pair of muons at the International Linear Collider (ILC).
This channel provides an opportunity to measure directly the Yukawa coupling between the Higgs boson and a second generation fermion.
The analysis is very challenging because the branching ratio $\mathrm{BR}(h\to\mu^+\mu^-)$ is predicted to be only $2.2 \times 10^{-4}$ in the Standard Model (SM). 

A MC study based on full simulation of the ILD detector~\cite{Behnke:2013lya} will be presented here.
A preliminary version of this analysis has been reported already at LCWS2017~\cite{LCWS2017} and a full publication of the final results is in preparation.
Thus, in this contribution, we will only summarize briefly the updates from~\cite{LCWS2017}.

\section{Analysis}
The analysis comprises 8 channels in total: two final states ($e^+ e^- \to q\overline{q}h$ and $e^+ e^- \to \nu \overline{\nu} h$) are studied for four  different data sets (at $\sqrt{s} =$ 250\,GeV and 500\,GeV, with two beam polarization configurations each).
In all cases, the full SM background is included.
At each center-of-mass energy, the integrated luminosity is shared between the four possible beam helicity configurations according to the official ILC running scenarios~\cite{Barklow:2015tja}.
In particular for the 250\,GeV case, the H20-staged scenario has been adopted~\cite{Fujii:2017vwa}.

The event selection is structured in the same way in all channels.
First, a pair of well-measured, prompt, oppositely charged muons is selected as the $h \to \mu ^+ \mu ^-$ candidate.
Then, the rest of the event is subject to a channel-specific analysis.
Various pre-selection cuts are applied to select signal-like events and suppress SM background processes, followed by a final selection based on a gradient boosted decision tree (BDTG).
A full description of the selection can be found in~\cite{LCWS2017}.

Finally, a toy Monte-Carlo (MC) technique is applied to extract the expected precision on cross section times branching ratio in each of the 8 channels.
In the following we will focus on the updates on this toy MC step w.r.t.~\cite{LCWS2017}.

\subsection{Toy MC}
Due to the excellent momentum resolution of the ILD detector, the distribution of the invariant mass of muon pair $M_{\mu ^+ \mu ^-}$  is crucial input to the final separation of signal and background.
After the BDT score cut, the shape signal and background distributions of $M_{\mu ^+ \mu ^-}$ are parametrised by functions $f_S$ and $f_B$  and their normalisations are then obtained from a fit to the pseudo-data histogram.
In order to avoid limitations due to the finite number of fully simulated events available, in particular for the SM background, this fit is repeated many times on toy data generated from the original parametrisations.

\begin{figure}[t]
\centering
\includegraphics[width=9truecm]{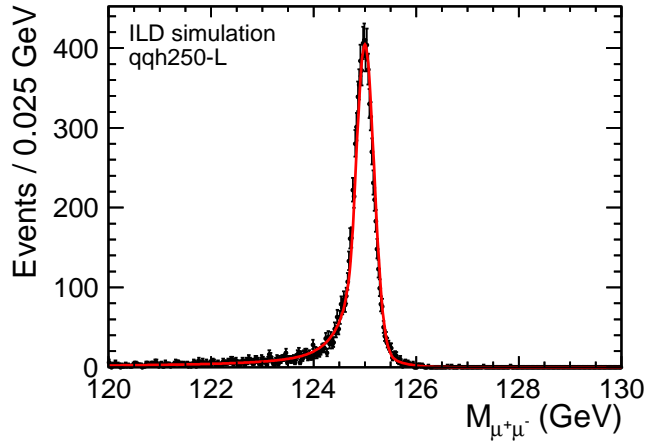}
\caption{Signal-only histogram as obtained from full simulation of the ILD detector fitted with the CBG function.
The normalisation is arbitrary.}
\label{fig:CBGfit}
\end{figure}

In~\cite{LCWS2017}, a normalized Gaussian was used to model  the shape of the signal distribution.
However, this approach was not fully satisfactory, e.g.:\ for modeling tails caused by final-state radiation of the muons.
Instead, a linear sum of a Crystal Ball function and a Gaussian (CBG) rendered the signal description much more reliable: $f_S = k \times \mathrm{Crystal \ Ball} + (1-k) \times \mathrm{Gaussian} \ (0 < k < 1)$.
An example fit to a signal-only histogram in arbitrary normalisation is shown in Figure~\ref{fig:CBGfit}.
The background modeling function $f_B$ has also been improved and is now a first order polynomial.

\begin{figure}[htb]
\centering
\includegraphics[width=14truecm]{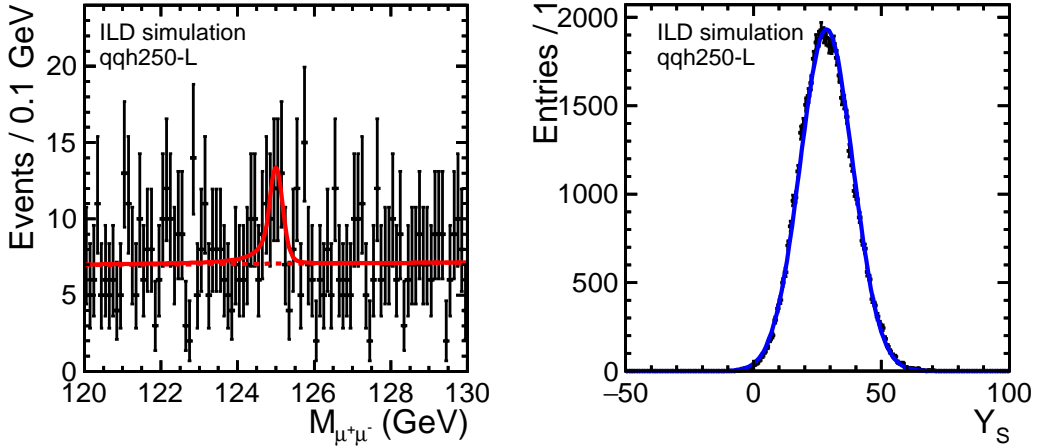}
\caption{Left: Example of one pseudo-experiment.
The solid red curve shows the fit result using the function $f$ as defined in the text, the dotted red curve shows the background component $Y_Bf_B$ only.
Right: $Y_S$ distribution after 50000 pseudo-experiments, with the Gaussian fit in blue.}
\label{fig:ToyMC}
\end{figure}

The parametrisations $f_S$ and $f_B$ are then used both as input for generating toy distributions, corresponding to a pseudo-experiment each.
In one pseudo-experiment, the number of signal (background) events are obtained from the number of full simulation events after the BDTG score cut $N_S \ (N_B)$, normalised according to luminosity and polarisation, plus a random Poisson fluctuation.
The pseudo-data are then subjected to an unbinned fit with the function $f \equiv Y_S f_S + Y_B f_B$, where $Y_S(Y_B)$ is the yield of signal(background).
Thereby, the $Y_B$ can be fixed, since at a lepton collider the SM background should be predicted to much higher precision than conceivable as statistical uncertainty for a rare signal like $h\to\mu^+\mu^-$.
Figure~\ref{fig:ToyMC} left shows an example of one pseudo-data set fitted with $f$.

This procedure is repeated 50000 times with different toy experiments.
A Gaussian function is fitted to the distribution of the 50000 resulting values for $Y_S$, as shown in the right panel of Fig.~\ref{fig:ToyMC}.
The final precision is calculated from the Gaussian width divided by the Gaussian mean.

The cut value on the BDTG score has been optimized for each channel using the full toy MC procedure: First, a cut on BDTG score is applied.
The resulting signal and background distributions are fitted with $f_S$ and $f_B$, respectively to determine their parameters.
Based on these parametrisations, the 50000 pseudo-experiments are performed.
Finally, the precision is evaluated from a Gaussian fit to the $Y_S$ distribution.
These four steps have been repeated for many different cuts on the BDTG score and the best cut is chosen for the analysis.

\subsection{Results}
\label{subsec:results}
The precisions on cross section times branching ratio $\sigma \times \mathrm{BR}(h \to \mu ^+ \mu ^-)$ obtained in the eight individual channels are summarized in Table~\ref{tab:precision}.

\begin{table}[htb]
\centering
\caption{Precision on $\sigma \times \mathrm{BR}(h \to \mu ^+ \mu ^-)$ for each channel.
The symbols L and R denote $\mathcal{P}(e^-.e^+)=(-80\%,+30\%)$ and $\mathcal{P}(e^-.e^+)=(+80\%,-30\%)$, respectively. }
\begin{tabular}{ccc}
\hline
$\sqrt{s} =$ 250\,GeV & $q\overline{q}h$ & $\nu \overline{\nu} h$ \\
\hline
L & 36.2{\%} & 122.4{\%}  \\
R & 38.0{\%} & 105.1{\%}  \\
\hline
$\sqrt{s} =$ 500\,GeV & $q\overline{q}h$ & $\nu \overline{\nu} h$ \\
\hline
L & 43.8{\%} & 37.9{\%}  \\
R & 54.2{\%} & 108.8{\%}  \\
\hline
\end{tabular}
\label{tab:precision}
\end{table}

The combined result for 2\,ab$^{-1}$ at $\sqrt{s} = $ 250\,GeV is 24.9{\%}, improving to 17.5{\%} when combined with 4\,ab$^{-1}$ at $\sqrt{s} =$ 500\,GeV\footnote{The quoted integrated luminosities comprise the data foreseen to be taken with like-sign polarisation configurations~\cite{Barklow:2015tja}.}.
These results are somewhat worse than the prospects for the High-Luminosity Large Hadron Collider (HL-LHC), where a precision on the signal strength of $h \to \mu ^+ \mu ^-$ between 10{\%} and 13{\%} is expected~\cite{HLLHCATLAS, HLLHCCMS}.
This difference mainly comes from the number of signal events; $O(10^4)$ $h \to \mu ^+ \mu ^-$ events will be produced at the HL-LHC while in total only about $\sim$ 200 signal events will be produced at the ILC.
In absence of backgrounds and with a perfect detector, these signal events would yield a combined precision on $\sigma \times \mathrm{BR}(h \to \mu ^+ \mu ^-)$ of about 10{\%} at $\sqrt{s} =$ 250\,GeV.
This improves to 7{\%} when adding the $\sqrt{s} =$ 500\,GeV data.
The full simulation results are about a factor of 2.5 far away from this ideal case.
There are several reasons: First, the typical signal selection efficiency is $\sim 50{\%}$.
At the same time, some irreducible backgrounds remain.
For instance events originating from $e^+ e^- \to W^+W^- \to \nu \overline{\nu}\mu ^+ \mu ^-$ are very hard to distinguish from the $\nu \overline{\nu} h$ signal.
The main discrimination power against this background is the different invariant mass distribution.
Therefore, a more narrow signal mass peak directly reduces the effective number of events from the rather flat background distribution which lie directly beneath the peak.
The role of the momentum resolution will therefore be discussed in more detailed in Sec.~\ref{sec:momres}.

It should also be noted that in the $\nu \overline{\nu} h$ process, especially at $\sqrt{s} =$ 500\,GeV, two signal processes  interfere with each other: the $Zh$ process with $Z \to \nu \overline{\nu}$ and the $WW$-fusion process.
The relative contributions of these production modes will be fixed to the percent-level or better from other Higgs decay modes, e.g.:\ $h \to b \overline{b}$, and can be used to convert the cross section times branching ratio measurement into a measurement of $\mathrm{BR}(h \to \mu^+\mu^-)$.
With the help of the total $Zh$ cross section determined with the recoil method, the absolute $h\mu\mu$ Yukawa coupling can be extracted.

\section{Impact of the Transverse Momentum Resolution}
\label{sec:momres}
Specifically for this analysis, the variable $M_{\mu ^+ \mu ^-}$ is the most important.
Thus, the transverse momentum resolution $\sigma_{1/P_t}$ has a crucial role for this analysis, because the muons are mainly measured by the tracking system of ILD.
The design goal for the ILD tracking system is an asymptotic transverse momentum resolution of $\sigma_{1/P_t} = 2 \times 10^{-5}$\,GeV$^{-1}$~\cite{Behnke:2013lya}.

The dependence of our result on the asymptotic value of the momentum resolution has been studied by smearing the MC truth momentum with a Gaussian random number.
All other quantities in the event are taken from full simulation as before.
Since this analysis is dominated by high energetic, central tracks, the asymptotic value momentum resolution has been used for all muons, the dependence of the resolution on the momentum itself and on the polar angle have not been considered.

The momentum resolution has been varied in a range from $10^{-3}$ to $10^{-6}$ (GeV$^{-1}$).
The same analysis procedure including the toy MC has been performed for each of these assumed resolutions, replacing only the momenta of the muons (and derived quantities) by the smeared truth values.
The background is kept unchanged from the full simulation study, since its invariant mass distribution after BDTG score cut is almost flat as can be seen e.g.: in Fig~\ref{fig:ToyMC}, and applying Gaussian smearing to a flat distribution will also make similar flat distribution.

\begin{figure}[t]
\centering
\includegraphics[width=12truecm]{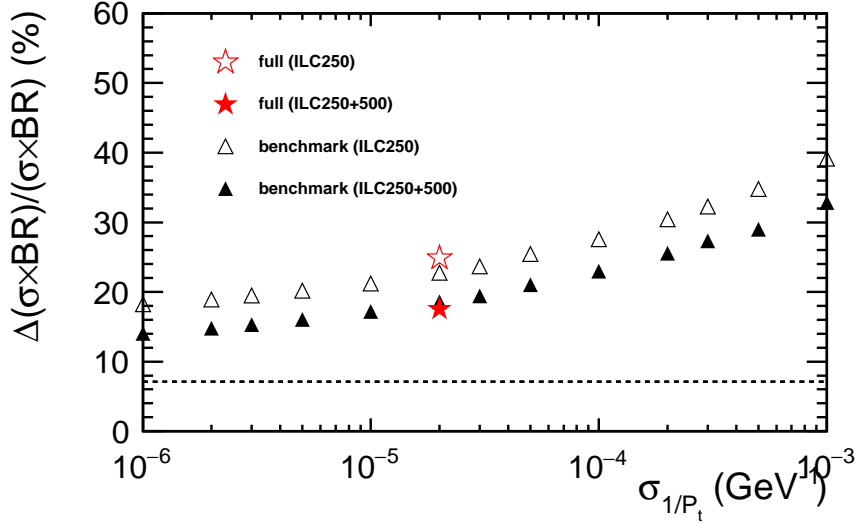}
\caption{Precision on $\sigma \times \mathrm{BR}(h \to \mu ^+ \mu ^-)$ as a function of the asymptotic transverse momentum resolution $\sigma _{1/P_t}$, together with the full simulation results indicated by the star symbols. ILC250(ILC250+500) means the combined results using $\sqrt{s} =$ 250 GeV results (using $\sqrt{s} =$ 250 GeV and 500 GeV results).
The bottom dotted-line shows the "theoretical limit" of 7{\%} which is explained in Sec~\ref{subsec:results}.}
\label{fig:momres}
\end{figure}

Figure~\ref{fig:momres} shows the precision on $\sigma \times \mathrm{BR}(h \to \mu ^+ \mu ^-)$ as a function of the transverse momentum resolution $\sigma _{1/P_t}$.
As expected, the better resolutions give better results.
In this direction, the potential gain is very limited: Even with an astronomical improvement in terms of detector technology, e.g.:\ by a factor 10 to $2 \times 10^{-6}$, the relative improvement from the full simulation result is only about 20{\%}.
However, in other direction, when the momentum resolution deteriorates by a factor of 10 to $2 \times 10^{-4}$, the precision will be significantly worse by nearly a factor of 1.5.
Therefore, we can conclude that it is very important for this analysis to reach the ILD design goal for the momentum resolution. 

\section{Summary}
We have studied the prospects of measuring decay of the Higgs boson into a pair of muons at the ILC at center-of-mass energies of 250\,GeV and 500\,GeV in full detector simulation of the ILD detector concept. The precision on $\sigma \times \mathrm{BR}(h \to \mu ^+ \mu ^-)$ can reach 24.9{\%} at the initial ILC stage, and improve to 17.5{\%} when combined with the $\sqrt{s} =$ 500\,GeV program.
The results have been evaluated for different assumptions on the transverse momentum resolution, and we conclude that it is important to achieve the design goal of $\sigma_{1/P_t} = 2 \times 10^{-5}$\,GeV$^{-1}$.

\section*{Acknowledgements}
We would like to thank the LCC generator working group and the ILD software working group for providing the simulation and reconstruction tools and producing the Monte Carlo samples used in this study.
This work has benefited from computing services provided by the ILC Virtual Organization, supported by the national resource providers of the EGI Federation and the Open Science GRID.
We thankfully acknowledge the support by the Deutsche Forschungsgemeinschaft (DFG) through the Collaborative Research Centre SFB 676 Particles, Strings and the Early Universe, project B1.

\end{document}